\documentclass[letter]{aa} % for the letters 
\usepackage{dcolumn}

\usepackage{graphicx}
\usepackage{txfonts}
\makeatletter
 \def\hlinewd#1{%
   \noalign{\ifnum0=`}\fi\hrule \@height #1 \futurelet
    \reserved@a\@xhline}
\makeatother

\begin{document}

   \title{Dust reverberation-mapping of the Seyfert 1 galaxy WPVS48}

   \titlerunning{Dust reverberation mapping of WPVS48}

	\author{
          F. Pozo Nu\~nez 
          \inst{1} 
          \and
          M. Haas          
          \inst{1}
          \and
          R. Chini
          \inst{1,2}
          \and
          M. Ramolla           
          \inst{1}
          \and
          C. Westhues
          \inst{1}
          \and    
          K. Steenbrugge
          \inst{2,3}		
          \and
          L. Kaderhandt
          \inst{1}
          \and   
          H. Drass
          \inst{1}
          \and  
          R. Lemke
          \inst{1}
          \and
          M. Murphy
          \inst{4}		
	}
	\institute{
          Astronomisches Institut, Ruhr--Universit\"at Bochum,
	  Universit\"atsstra{\ss}e 150, 44801 Bochum, Germany
	  \and
          Instituto de Astronomia, Universidad Cat\'{o}lica del
          Norte, Avenida Angamos 0610, Casilla
          1280 Antofagasta, Chile
          \and
          Department of Physics, 
          University of Oxford, 
          Keble Road,
          Oxford OX1 3RH, UK
          \and
  Departamento de F\'{i}sica, Universidad Cat\'{o}lica del
  Norte, Avenida Angamos 0610, Casilla
  1280 Antofagasta, Chile
        }
        
	\authorrunning{F. Pozo Nu\~nez et al.}

	\date{Received ; accepted}

% \abstract{}{}{}{}{} 
% 5 {} token are mandatory
 
\abstract{
  Using robotic telescopes of the 
  Universit\"atssternwarte Bochum near Cerro Armazones in Chile, 
  we monitored the $z=0.0377$ Seyfert-1 galaxy WPVS48 
  (2MASX J09594263-3112581) in the optical ($B$ and $R$) and 
  near-infrared (NIR, $J$ and $K_{s}$) with a cadence of two days.
  The light curves show unprecedented variability details.
  The NIR variation features of WPVS48 are consistent with
  the corresponding optical variations, but  
  the features appear sharper in the NIR than in the optical, 
  suggesting that the optical photons undergo multiple scatterings. 
  The $J$ and $K_{s}$ emission, tracing the hot ($\sim$1600 K) 
  dust echo, lags the $B$ and $R$ variations by on average 
  $\tau = 64 \pm 4$ days and $71 \pm 5$ days, respectively (restframe). 
  WPVS48 lies on the known $\tau - M_{V}$ relationship. 
  However, the observed lag 
  $\tau$ is about three times shorter than expected from 
  the dust sublimation radius $r_{sub}$ inferred 
  from the optical-UV luminosity, and explanations 
  for this common discrepancy are searched for. 
  The sharp NIR echos argue for a face-on torus geometry 
  and allow us to put forward two potential scenarios: 
  1) as previously proposed, in the equatorial plane of the
  accretion disk the inner region of the torus is flattened
  and may come closer to the accretion disk.
  2) The dust torus with inner radius $r_{sub}$ 
  is geometrically and optically thick, 
  so that the observer only sees the facing rim of 
  the torus wall, which lies closer to the observer 
  than the torus equatorial plane and 
  therefore leads to an observed foreshortened lag. 
  Both scenarios are able to explain the factor three 
  discrepancy between $\tau$ and $r_{sub}$. 
  Longer-wavelength dust reverberation data 
  might enable one to distinguish between the scenarios.
}    
		\maketitle
%
%________________________________________________________________

\section{Introduction}

In the standard paradigm, active galactic nuclei (AGNs) are composed
of a central black hole with an accretion disk (AD). A broad line region
(BLR) surrounds the AD, and farther out there is a (clumpy) molecular dust
torus (\citealt{1993ARA&A..31..473A}). These components cannot be spatially resolved
by conventional imaging, but reverberation mapping (RM, \citealt{1982ApJ...255..419B}) can separate them.  
In RM, continuum brightness variations of the AD are echoed by the BLR emission lines and by the torus dust emission.
The time delay $\tau$ gives the characteristic size ($R = \tau \cdot c$, where $c$ is the speed of light) and in favorable cases the geometry of the respective regions. 

The geometry of AGN dust distributions is still under debate. Dust RM
of Seyfert 1 galaxies (\citealt{2004MNRAS.350.1049G},  
\citealt{2004ApJ...600L..35M}, \citealt{2006ApJ...639...46S}) revealed a relation 
$R_{\rm in} = \tau \cdot c \propto L_{UV}^{0.5} $ between the 
inner radius $R_{\rm in}$ of the dust torus and the UV luminosity
$L_{\rm UV}$ (\citealt{1999AstL...25..483O}, \citealt{2007A&A...476..713K}). For a
bagel-shaped dust torus, however, the observed  $R_{\rm in}$ is
about a factor three smaller  
than the sublimation radius $R_{\rm sub}$ predicted from the dust
sublimation temperature $T_{\rm sub} \sim 1600$~K (\citealt{1992ApJ...400..502B}). Some modified dust geometries involve BLR associated dust due
to winds/outflows from the accretion disk (\citealt{1994ApJ...434..446K},
\citealt{2006ApJ...648L.101E}, \citealt{2011A&A...525L...8C}). Non isotropic 
illumination of the torus by the AD has been proposed as well
(\citealt{2010ApJ...724L.183K}), allowing the dust to come closer in the AD's
equatorial plane without being sublimated. On the other hand, the
3~$\mu$m emission bump in the spectral energy distribution of type 1
AGN might be attributed to a puffed-up inner rim of the torus (\citealt{2007ApJ...661...52K}). 

WPVS48 is a nearby Seyfert-1 galaxy ($z=0.0377$) located at a distance
of 161 Mpc (\citealt{2010A&A...518A..10V}). Single-epoch UBVRI
multi-aperture photometry by \cite{1997MNRAS.292..273W} showed evidence of
a luminous nucleus ($V=14.78$). 
Here we present results for the dust
torus geometry of WPVS48. Based on a dust RM
campaign, we determine inner size of the dust torus and the
host-subtracted AGN luminosity. The sharp near-infrared (NIR) variation in WPVS48
allow us to favor two geometric models that are able to explain the
factor of three discrepancy between $R_{\rm in} = \tau \cdot c$ and 
$R_{\rm sub}$ in Seyfert 1 galaxies. 

%__________________________________________________________________

%------------------------------------------------------------------------------
%
\begin{figure*}
\includegraphics[width=0.67\columnwidth]{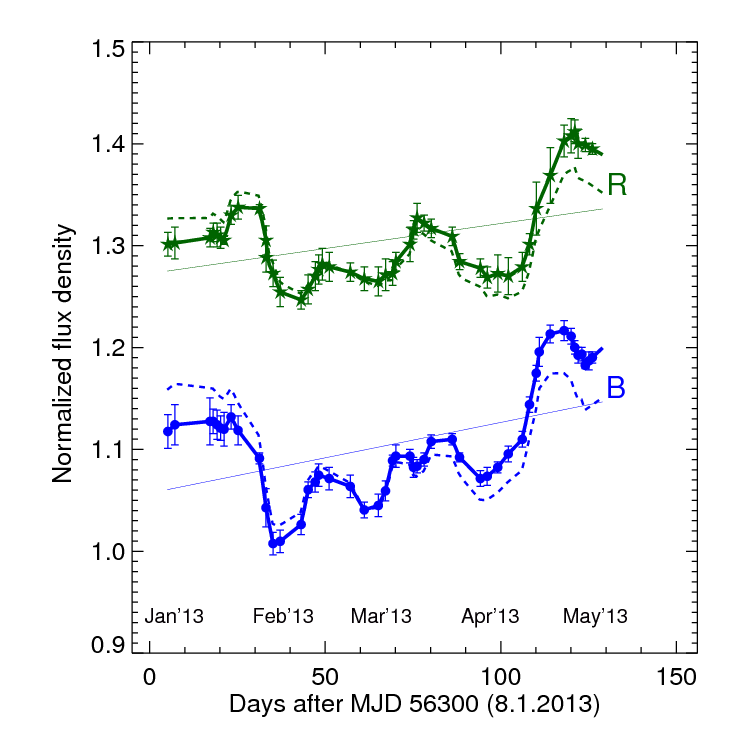} 
\includegraphics[width=0.67\columnwidth]{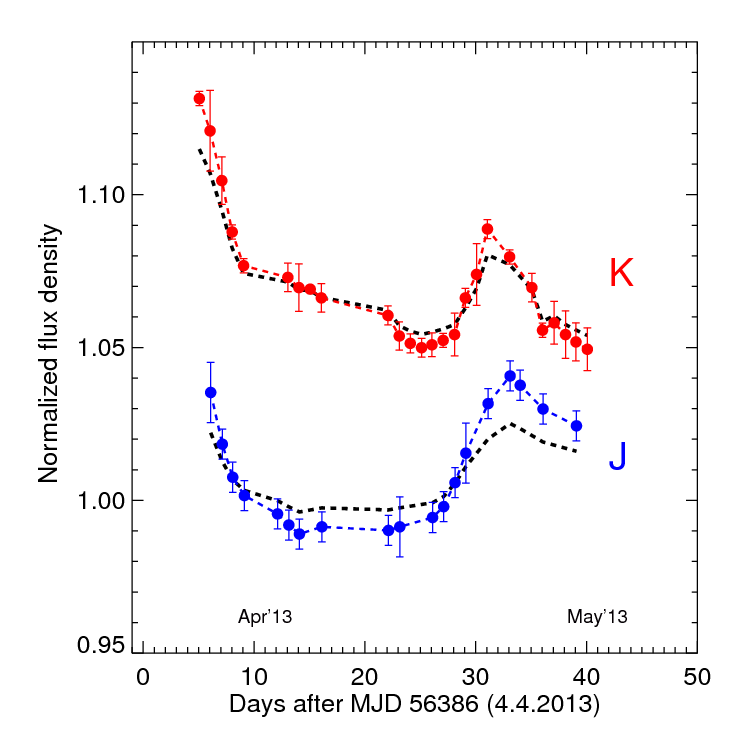} 
\includegraphics[width=0.67\columnwidth]{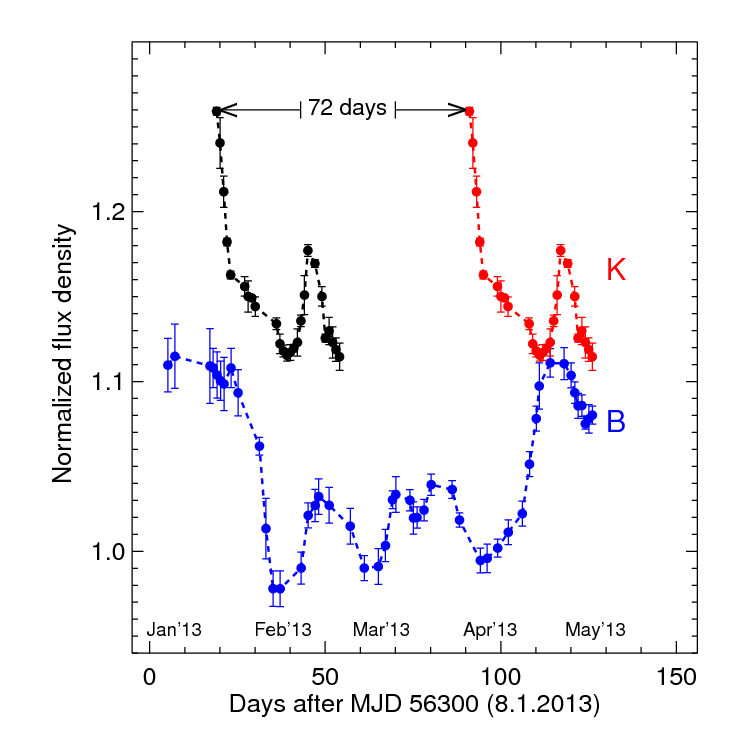} 
%       \vspace*{5mm} 
  \caption{
    {\it Left:} 
    $B$- and $R$-band light curves obtained in 2013 January-May. 
    The $R$ light curve is vertically shifted for clarity.  
    The straight lines indicate the long-term brightness trends.
    The dotted lines show the detrended light curves.
    {\it Middle:} 
    $J$- and $K$-band light curves obtained in 2013 April-May. 
    The $K$ light curve is vertically shifted for clarity. 
    The dotted light curves depict the state before host galaxy subtraction.
    {\it Right:} 
    $B$- and $K$-band light curves (blue and red) after host
    galaxy subtraction.  
    The $K$-band variation pattern (red), when shifted back by
    $\sim$72 days (black), 
    largely corresponds
    to the $B$-band variation pattern in February 2013; see text for details.
  }
   \label{fig1}
\end{figure*}
%
%----------------------------------------------------------------------------- 

\section{Observations and data}

%\subsection{Optical monitoring}

%__________________________________________________________________

The optical Johnson broad-band $B$ ($4330$\,\AA) and $R$ ($7000$\,\AA)
observations were carried out between January 12 and May 13 in 2013,
with the robotic 25\,cm Berlin Exoplanet Search Telescope-II
(BEST-II), located at the Universit\"atssternwarte Bochum, near Cerro
Armazones,
Chile\footnote{http://www.astro.ruhr-uni-bochum.de/astro/oca/}. As in
\cite{2013A&A...552A...1P}, we performed standard data reduction
including corrections for bias, dark current, flatfield, astrometry and astrometric
distortion before combining 
the nine dithered images of each night and filter. 
A 7$\farcs$5 diameter aperture was used to
extract the photometry and to create flux-normalized light curves
relative to 17 non variable stars located in the same images, within
30$\arcmin$ around the AGN, and of similar brightness as the
AGN. Absolute calibration was performed using standard reference stars
from \cite{2009AJ....137.4186L} observed on the same nights as the AGN.
We also corrected for atmospheric (\citealt{2011A&A...527A..91P}) and Galactic
foreground extinction (\citealt{2011ApJ...737..103S}).  

%__________________________________________________________________

%\subsection{Near-infrared monitoring}

After observing the pronounced optical variations in February
2013, we performed the NIR $J$ and $K_{s}$ 
(hereafter denoted as $K$)
observations between April 10 and May 13 in 2013 using the 0.8\,m
Infrared Imaging System (IRIS) telescope (\citealt{2010SPIE.7735E..44H}) at the
Universit\"atssternwarte Bochum. Images were obtained  
and reduced in the standard manner. 
Photometry was extracted using a 7$\farcs$0 diameter
aperture, slightly smaller than for the optical light curves, because
the seeing in the NIR images is slightly better and smaller pixel sizes were used. 
Light curves were calculated relative to six non variable stars
located in the same field that have a similar brightness as the AGN. The
photometric calibration was achieved using four high-quality flag
(AAA) 2MASS stars in the same field as the
AGN. Absolute fluxes were corrected for Galactic foreground
extinction (\citealt{2011ApJ...737..103S}). 

\section{Results and discussion}

\subsection{Optical and infrared light curves}

%__________________________________________________________________

The light curves of WPVS48 are shown in Fig~\ref{fig1}. A summary of the photometric
results and the fluxes in all bands is listed in Table\,1. The light curves are 
published in electronic format. The two
optical and the two NIR light curves show a similar variability
behavior.  
The optical light curves exhibit a long-term trend that we corrected for,
following \cite{1999PASP..111.1347W} and \cite{2010ApJ...721..715D}, to avoid
a potential bias in the cross-correlation analysis. 
The NIR light curves are too short for detrending.

\begin{figure}
\includegraphics[width=0.475\columnwidth]{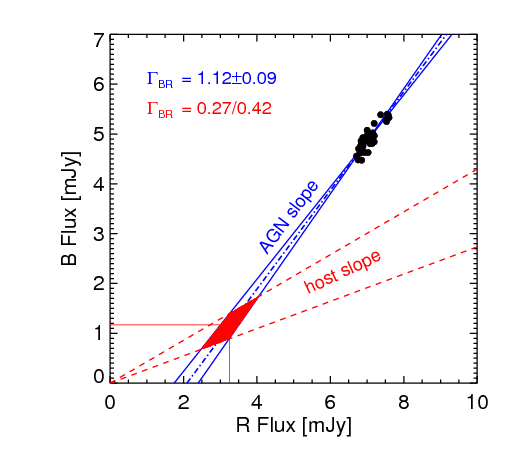}
\includegraphics[width=0.475\columnwidth]{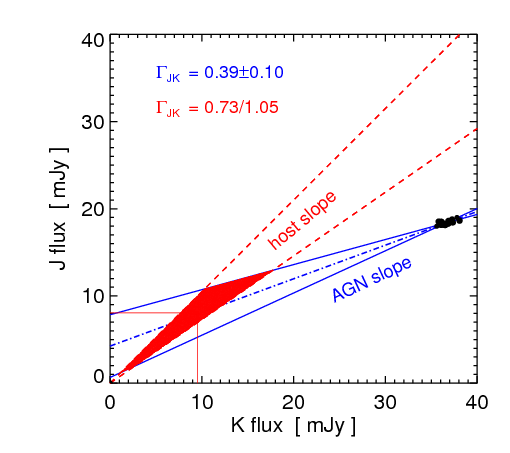} 
  \caption{Flux variation gradient diagrams of WPVS48 
    in the optical (left) and NIR (right). 
    The data are represented by the black dots.
    The solid blue lines represent the
    bisector fit, yielding the range of the AGN slopes. 
    The dashed red lines indicate the range of host slopes determined
    in the optical by \cite{2010ApJ...711..461S} and the NIR by \cite{2006ApJ...639...46S}.  
    The intersection between the host galaxy and AGN slope (red area)
    gives the host galaxy flux in the respective bands.
  }
   \label{fig2}
\end{figure}

Using the flux variation gradient (FVG) method (\citealt{1981AcA....31..293C};
\citealt{1992MNRAS.257..659W}, \citealt{2004MNRAS.350.1049G}, \citealt{2010ApJ...711..461S}), we determined
the host galaxy contribution to the light curves (Fig.~\ref{fig2}). In
brief, it was observationally established that the optical and
near-infrared flux ratios (B/R and J/K) of the variable component remain
constant with 
time, allowing us to separate the AGN flux 
through the use of a well-defined range of host
galaxy colors (\citealt{2004MNRAS.350.1049G}, \citealt{2010ApJ...711..461S}). More details on
the FVG method are given in \cite{2012A&A...545A..84P}.
As noted by \cite{2004MNRAS.350.1049G}, 
the $K$-band light curve may be contaminated by slower-varying 
longer-wavelength ($L$ band) emission, and we address this effect in
Sect.\,\ref{section_torus_geometry}.
Power-law extrapolation of the AGN $B$ and $R$ fluxes 
(presumably from the AD) to longer wavelengths
shows that the contribution of the AD to the $J$ and $K$ fluxes 
is negligible ($<$10\%).
  
\begin{figure}
\includegraphics[width=0.475\columnwidth]{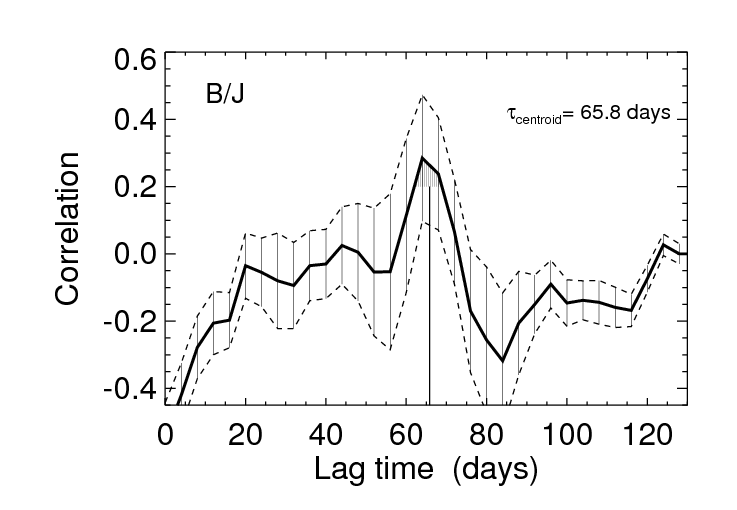} 
\includegraphics[width=0.475\columnwidth]{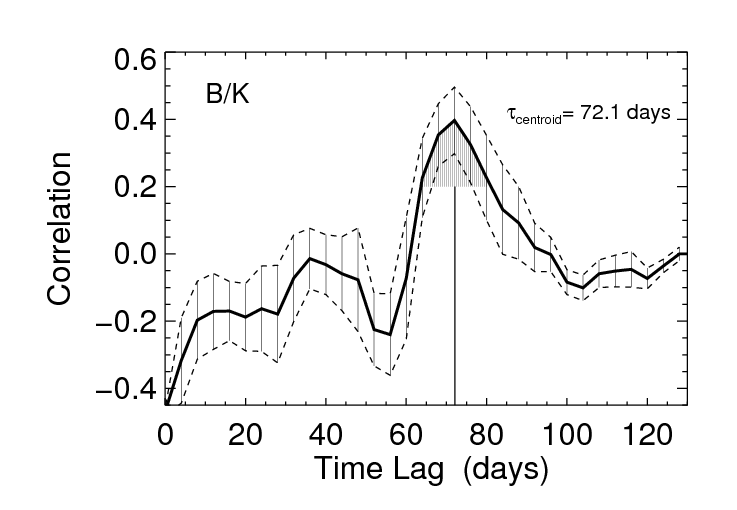} 
  \caption{
    Discrete correlation function of the $B$ and $J$ and the $B$ and $K$
    light curves. The dotted lines indicate the error range ($\pm
    1\sigma$) around the cross correlation. The black shaded area
    marks the range 
    used to calculate the centroid of the lag (vertical line).
  }
   \label{fig3}
\end{figure}

We correlated both the $B$- and $K$-band light curves and the $B$- and
$J$-band light curves using the discrete
correlation function (DCF, \citealt{1988ApJ...333..646E}). The DCF centroid
yields a time delay $\tau = 65.8$ days of $B/J$ and 
$\tau = 72.1$ days of $B/K$
 (Fig~\ref{fig3}). The
uncertainties of $\tau$ were calculated using the flux randomization
and random subset-selection method (FR and RSS, \citealt{2004ApJ...613..682P}). 
The median of this procedure yields
$\tau_{cent}$ = 66.5 $^{+3.8}_{-4.1}$ days and $\tau_{cent}$ = 73.5 $^{+4.4}_{-5.2}$ days for $B/J$ and $B/K$, respectively. Correcting for the time
delation factor, we obtain a rest frame time-delay $\tau_{rest} = 64.1
\pm 3.81$ days and $\tau_{rest} = 70.8 \pm 4.63$ days for $B/J$ and $B/K$, respectively.

When the $K$ light curve is shifted back by 72 days 
(black curve in Fig~\ref{fig1}, right), the variation
features (the steep decline, the valley, and the subsequent peak) 
are roughly consistent with the $B$ features.
However, the $K$ decline precedes the  $B$ decline, 
the subsequent peak is sharper in $K$ than in $B$,
and the time span of the $K$ valley (between the decline and the
subsequent peak) is longer than for $B$.
The same holds for the $R$ and $J$ curves, respectively.
An explanation 
(private communication by Ralf
Siebenmorgen and Endrik Kr\"ugel; see also \citealt{2009A&A...493..385K}) 
might be that on the way to
the observer a significant fraction of the optical 
photons are scattered multiple times in the kpc-scale 
narrow-line region. 
This leads to a longer traveling path to the observer, hence a 
delayed arrival. The net effect is that we observe the AD variability 
convolved with an asymmetric delay kernel. 
The longer-wavelength NIR
photons may be scattered in the torus, but with small path-length
changes, and they are only weakly affected by scattering in the
NLR, so that their
observed variation features remain sharp, even sharper than the
optical features. 

One might expect that AD brightening leads to  
dust destruction and the receding of the dust sublimation front, and vice
versa.  
Then after a bright NIR light curve plateau a decline should be delayed 
and a rise after a low state should occur earlier, 
but the variations of the current NIR data are too small to allow an 
identification of such signatures.

\begin{figure}
  \centering
   \includegraphics[width=0.9\columnwidth,clip=true]{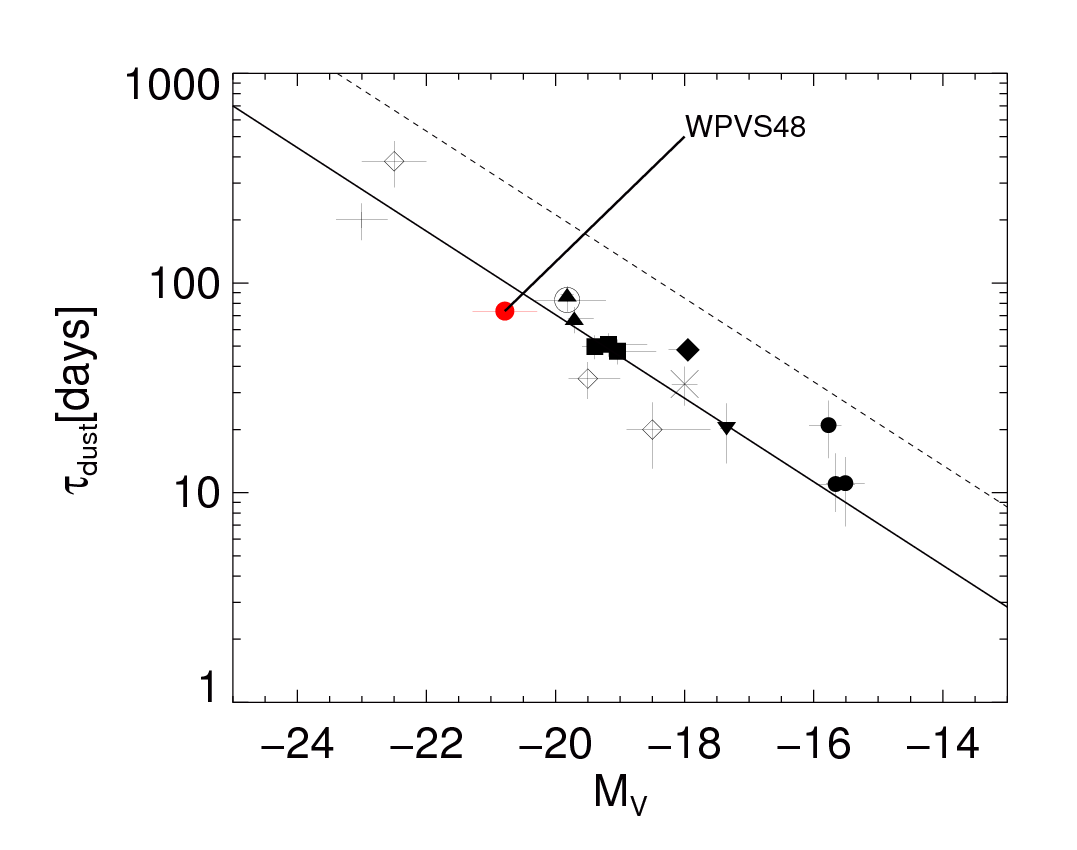}
  \caption{
    Lag -- luminosity relationship based on the data of \cite{2006ApJ...639...46S}. The symbols for individual Seyfert 1 galaxies are the
    same as in \cite{2006ApJ...639...46S}. The solid line is the best-fit
    regression from \cite{2006ApJ...639...46S}. WPVS48 (red dot) lies
    close to the regression line. The dashed line indicates the dust
    sublimation radius $r_{sub}$ expected at a given
    nuclear luminosity $M_{V}$ (from \citealt{2007A&A...476..713K}).
  }
  \label{fig4}
\end{figure}

Dust-reverberation studies of ten Seyfert 1 galaxies have shown that
the lag $\tau$ of the dust $K$-band emission is proportional to the
square root of the optical luminosity ($\tau \propto L^{0.5}$;
\citealt{2006ApJ...639...46S} and references therein). 
We interpolated $M_{V}$ of WPVS48 from our $B$ and $R$
data.
Figure~\ref{fig4} shows
its position in the $\tau - M_{V}$ diagram close to the
regression line.

\subsection{Inner geometry of the dust torus}
\label{section_torus_geometry}

Photometric H$\alpha$ reverberation-mapping of
WPVS48 revealed that the echo of the H$\alpha$ BLR
has a mean lag of 25 days (Pozo Nu\~nez et al. in prep). In
consequence, because of the different lags of H$\alpha$ and dust echo,
BLR-related dust probably plays a minor role.  

If the dust distribution is spread far along the line-of-sight, 
one would expect that the echo is smeared out in time. 
Therefore
the sharp dust echo of
WPVS48 
argues in favor of a face-on torus geometry.
We focused on the basic geometry of the overall absorbing dust, 
and considered smooth dust distributions.
The topic of a clumpy dust structure is left for future modeling.

A general puzzle from dust RM is that
with reasonable assumptions for graphite grains of a 
sublimation temperature $\sim$1600\,K and size 
0.05$\mu$m in radius (\citealt{1992ApJ...400..502B}) and for a bagel-shaped torus, 
the 
inner radius $R_{\rm in} = \tau \cdot c$ is about three
times smaller than  
the dust sublimation radius $r_{sub}$ 
inferred from the optical-UV luminosity (\citealt{2007A&A...476..713K}). 

\begin{figure}
\includegraphics[width=\columnwidth]{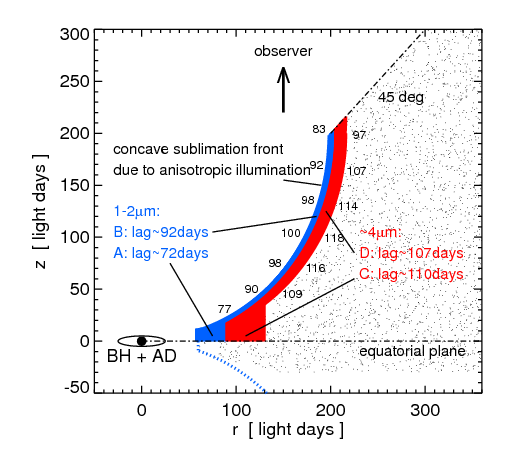} 
  \caption{Schematic illustration of the dust torus model by Kawaguchi
    \& Mori.  
    Blue denotes hot $T \sim 1600 K$ dust visible at 1-2 $\mu m$, red denotes
    cooler $T \sim 800 K$ dust visible at $\sim$4 $\mu m$.
    The single black numbers left and right of the rim 
    give the lags along the torus rim at 1-2$\mu$m and at $\sim$4$\mu$m, 
    respectively. 
    Note the foreshortening of the lags at the high-altitude 
    part of the torus rim.
    The mean lags of 
    the equatorial zones (A and C) and the entire 
    2$\pi$ integrated torus rim (B and
    D) are given, adopting a constant dust density distribution
    along the rim.
  }
   \label{fig5}
\end{figure}

To solve this puzzle, 
Kawaguchi \& Mori (2010, hence 
KM2010, 2011) have considered a model where the 
accretion disk emits relatively little radiation in its equatorial plane. 
The anisotropic illumination allows the torus inner region in the 
equatorial plane to approach to the central black hole and shapes
the inner torus wall so that it is 
concave from the observers view, 
with the height $z$ above the equatorial plane increasing radially
with $r$ to a covering half-angle of $\sim$45$^\circ$.
In this picture, illustrated in Fig.~\ref{fig5},
the hot dust emission comes from a geometrically 
thin inner dust torus with a remarkably small dust covering 
angle, and along the dust sublimation rim. 

Specifically for WPVS48, 
the dust may approach in the AD's equatorial plane, 
without sublimation, and the hot dust will produce a 
sharp echo at mean lag $\sim$72 days, but the long 
inner torus rim will add a smeared-out echo of a longer lag
ranging between $80d > \tau < 100d$ 
(Fig.~\ref{fig5}).
The bulk of the dust echo of WPVS48 
is as sharp as the triggering continuum variations in the $J$ and $K$
light curves with a rapid decline in $J$ and a two-step (rapid and
subsequent shallow) decline in $K$
(Fig.\,\ref{fig1}, middle). In the cross
correlation $B$/$K$ a faint tail clearly extends to $\sim$100 days, 
but not in $B$/$J$ (Fig.\,\ref{fig3}). 
Therefore the current data of WPVS48 appear to
be 
not fully consistent with the expectations from the KM2010 model: 
Firstly, if the dust sublimation front is shaped by the anisotropic
illumination from the AD alone, as proposed by KM2010, one would expect that
the hottest $J$- emitting dust exists along the entire rim and that its
signatures seen in the $K$ light curve and the $B$/$K$ cross
correlation are also seen in $J$. 
Secondly, this interpretation of the slow shallow $K$ decline 
leaves no room for 
$K$-band variations that are associated in part with longer-wavelength $L$-band
flux (\citealt{2004MNRAS.350.1049G}).
Thirdly, while, in addition to the radiation fields,
AGN winds might help to shape the torus rim, 
the geometry of the AD 
(thin disk vs. disk becoming thicker with increasing radius) 
is unknown, and it is not clear how strong the
anisotropy actually is (e.g. \citealt{2013Natur.495..165A} and references therein).

\begin{figure}
\includegraphics[width=\columnwidth]{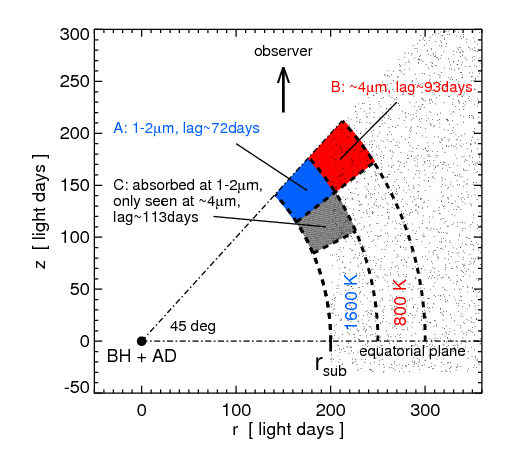} 
  \caption{Schematic illustration of a torus with constant
    sublimation radius for inclinations $0^\circ < i < 45^\circ$,
    optically thick at 1-2\,$\mu m$. 
    Zone A (blue) marks hot $T \sim 1600 K$ 
    dust visible at 1-2\,$\mu m$, zone B (red)
    cooler $T \sim 800 K$  
    dust visible at $\sim$4 $\mu m$; zone C (black shaded) 
    markes hot dust that is
    heavily obscured ($A_K > 1$) in the observer's line of sight, so
    that it becomes dim at 1-2\,$\mu m$,
    but is clearly visible at $\sim$4\,$\mu m$.
    The mean lags for each zone are given.
  }
   \label{fig6}
\end{figure}

We here also considered an alternative scenario, 
illustrated in Fig.~\ref{fig6}.
In this picture, for simplicity, 
we assumed that the AD essentially radiates isotropically. 
The dust torus is geometrically thick 
with inner radius $r_{sub}$ and an opening half 
angle $\sim45^\circ$ and 
an optically thick at NIR wavelengths, 
so that the observer only sees the facing rim of 
the entire inner dust torus wall.
The striking feature is that the rim lies at height $z$, that is, closer 
to the observer than the equatorial plane. 
This $z$ contribution foreshortens the observed lag, namely 
$\tau \cdot c = (r^{2}+z^{2})^{0.5} - z \approx  r_{sub} - z$. 
If the rim lies at an angle $45^\circ$ from the equatorial plane,
one obtains
$z = cos(45^\circ) \cdot r_{sub}$, hence 
$\tau \cdot c = 0.3 \cdot r_{sub}$. 
This scenario can explain the factor three discrepancy between $\tau$ and
$r_{sub}$ as an observational foreshortening effect.
The $J$ and $K$ light curves are consistent with this
picture: the rapid decline in $J$ and $K$ arises from dust in
the blue zone of Fig.~\ref{fig6} 
and the subsequent shallow decline in the $K$ light curve
might be explained by dust in
the transition zone between the blue zone and the red and black zones,
which are not warm enough (red zone) or too much absorbed
(black zone) in $J$. 
Then the shallow $K$ decline would be associated with longer-wavelength $L$-band flux, noted in other sources by \cite{2004MNRAS.350.1049G}. 

Future dust RM including longer wavelengths
(e.g. at 3.6 and 4.5\,$\mu$m) may
provide  
clues to distinguish between the two models.
Firstly, the contribution of $T \sim 800 K$ cool dust,
visible at 4\,$\mu m$, to the $K$ band can be
constrained.
Secondly, for the Kawaguchi \& Mori model (Fig.\,\ref{fig5}), 
the expected 4\,$\mu m$ lags are similar along the rim (component D)
and in the equatorial plane (component C).  
On the other hand, for the geometrically and optically thick torus
model (Fig.\,\ref{fig6}), 
the cool dust (component B) 
and the hot dust with high extinction (component C), as
seen by the observer, might produce different lags of equal strength,
resulting in light curves with a two-step behavior.

\begin{acknowledgements}
 
  This work is supported by the
  Nordrhein-Westf\"alische Akademie der Wissenschaften und der K\"unste
  in the framework of the academy program of the Federal Republic of
  Germany and the state Nordrhein-Westfalen, by 
  Deutsche Forschungsgemeinschaft (DFG HA3555/12-1) 
  and by Deutsches Zentrum f\"ur Luft-und Raumfahrt (DLR 50\,OR\,1106).
  
  We thank our referee Ian Glass for helpful comments
  and careful review of the manuscript.
  We warmly thank Endrik Kr\"ugel and Ralf Siebenmorgen for discussions.
  The observations on Cerro Armazones benefitted
  from the care of the guardians Hector Labra, Gerardo Pino, Roberto Munoz, 
  and Francisco Arraya.
  This research has made use of SIMBAD and the NASA/IPAC
  Extragalactic Database (NED). %operated by JPL/CALTECH, 
%  and of the SIMBAD database. %, operated at CDS, Strasbourg, France.

\end{acknowledgements}

%------------------------------------------------------------------------------
%

\begin{table}
\begin{center}
\caption{
  Photometry range in mag and mean flux densities with errors in mJy.
  The values are as observed in 7$\farcs$5 ($B,R$) and 7$\farcs$0 ($J,K$) apertures.
}
\label{table3}
\begin{tabular}{@{}cccc}
\hline
$B$ & $R$ & $J$ & $K$ \\ 
\hline
14.74-14.94 & 13.93-14.06 & 12.33-12.36 & 10.61-10.66 \\ 
\hline
$fB$ & $fR$& $fJ$ & $fK$ \\
\hline
4.90$\pm$0.04 & 7.04$\pm$0.07 & 18.36$\pm$0.46 & 36.56$\pm$1.10 \\
\hline
\end{tabular}
\end{center}
\end{table}

\bibliographystyle{aa} 
\bibliography{wpv_dust}

\end{document}